# Intrinsic inhomogeneities in manganite thin films investigated with scanning tunneling spectroscopy

T. Becker, C. Streng, Y. Luo, V. Moshnyaga, B. Damaschke, N. Shannon[1] and K. Samwer

I. Physikalisches Institut, Universitaet Goettingen, Bunsenstr. 9, 37073 Goettingen, Germany

[1]Present address: Max Planck Institut für Physik komplexer Systeme, Dresden, Germany

**Abstract**

Thin films of $La_{0.7}Sr_{0.3}MnO_3$ on MgO show a metal insulator transition and colossal magnetoresistance. The shape of this transition can be explained by intrinsic spatial inhomogeneities, which give rise to a domain structure of conducting and insulating domains at the sub micrometer scale. These domains then undergo a percolation transition. The tunneling conductance and tunneling gap measured by scanning tunneling spectroscopy were used to distinguish and visualize these domains.

The properties of manganites were fist investigated more than fifty years ago by Jonker and van Santen [1]. Since the discovery of a colossal magnetoresistance (CMR) effect in manganite thin films by von Helmolt et al. in 1993 [2], interest in the magnetic structure and the metal insulator transition has been renewed [3-5].

Undoped $LaMnO_3$ is an antiferromagnetic insulator with perovskite structure. By substituting the trivalent La by a divalent material like Ca or Sr, valence mixing of $Mn^{3+}$ and $Mn^{4+}$ occurs, and electrons can hop between different Mn ions. This simple double exchange mechanism [6, 7] leads to a ferromagnetic ground state with metallic conduction. Additionally, an interaction with the lattice takes place via Jahn-Teller distortions around the $Mn^{3+}$ ion. The degrees of freedom of spin, charge and orbital open the field for a rich phase diagram with antiferromagnetic, ferromagnetic and paramagnetic phases which may be conducting or insulating [8, 9].

In samples near the doping range of 0.3 the metal to insulator transition (MIT) is accompanied by a maximum in the electrical resistance as a function of temperature, and a pronounced CMR effect around the Curie temperature. In a fully homogenous sample one would expect a sharp transition in the resistivity from the metallic ferromagnetic low temperature phase into the insulating paramagnetic high temperature phase at the Curie temperature. The measurements however show a broad transition, and the CMR does not saturate at high magnetic field. An explanation for these effects has been proposed in terms of intrinsic inhomogeneities, which can lead to the formation of insulating and conducting domains within a single sample. The unusual form of the MIT seen in transport measurements can then be explained in terms of a percolation transition [10-14].

At low temperatures the behavior of the sample is governed by the metallic double exchange phase. With increasing temperature electrically insulating regions occur. A further increase of the temperature leads to a growth of the insulating domains. In this temperature regime metallic and insulating domains coexist ('electronic phase separation') in a chemically homogeneous sample. This 'intrinsic' behavior can additionally be altered by external parameters like pressure, external magnetic field or doping. Here we focus on the electronic phase structure in the vicinity of the MIT.

The proposed inhomogeneities alter the local electronic and magnetic properties of the sample and should therefore be visible in scanning tunneling spectroscopy (STS) and magnetic force microscopy (MFM) measurements. Fäth et al. [15] investigated samples of $La_{0.7}Ca_{0.3}MnO_3$ (LCMO) near the Curie temperature $T_c$ and found spatial variations of the local electronic properties on a sub micrometer scale forming 'cloud' like regions of conducting, insulating and intermediate phases. To distinguish these regions they used the local differential conductance dI/dV at a voltage of 3V, and investigated the development of the metallic regions in an external magnetic field. In MFM-measurements the magnetic domain structure was investigated on the scale of some µm [16]. In the neighbourhood of $T_c$ the ferromagnetic domains change form and vanish at $T_c$ in agreement with macroscopic magnetization measurements. Recently, Renner et al. [17] were able to achieve atomic resolution with the STM on $Bi_{0.24}Ca_{0.76}MnO_3$ single crystals in insulating and metallic phases.

Here we present measurements of the local electronic properties of $La_{0.7}Sr_{0.3}MnO_3$ thin films with STS from 48K to room temperature under UHV conditions. In contrast to Fäth et al. [15] the local tunneling conductance was measured from the derivative of the V-I-characteristics at zero voltage. Using this technique we were able to visualize the domain structure and to investigate the tunneling conductance of the metallic regions as a function of temperature.

The $La_{0.7}Sr_{0.3}MnO_3$ (LSMO) film with a thickness of 50 nm was prepared by reactive sputter deposition from a LSMO target [18, 19]. Additionally, results from a 200 nm thick $La_{0.7}Ca_{0.3}MnO_3$ (LCMO) film which was deposited by a metalorganic aerosol deposition (MAD) method [20] are presented. MgO substrates were used in order to get epitaxial growth of the films [21].

After transfering the samples into the ultra high vacuum (UHV) chamber they were cooled down to low temperature (< 50K) in order to avoid oxygen loss by diffusion from the surface of the thin films. This procedure was performed because the oxygen loss leads to highly insulating film, for which STS and STM measurements were not reproducible.

The STM/STS measurements were made with a variable temperature STM (temperature range 16K to 300K) working in UHV [Omicron, Taunusstein, Germany] using cut PtIr tips. For topography measurements (Fig. 1) a voltage of 1V and a current of 0.3nA is used. I-V-characteristics were scanned at each $5^{th}$ data point of the topography. The I-V curve was scanned in a voltage range from -0.6 V to 0.6 V.

Using this technique two kinds of characteristics were obtained, a metallic one with a linear region at the Fermi energy and an insulating one with a plateau around V=0 typical for an insulating gap (Fig. 2a). From the metallic characteristics a local conductance was evaluated by the following procedure: The measured I-V characteristics were fitted using a $5^{th}$ order polynomial, and the derivative was calculated in order to determine the slope of the curve at the origin, which represents a measure of the local tunneling conductance. The derivative at V=0 can then be plotted to get a 'conductance map' of the sample. Moreover, it is possible to calculate the local gap width from the second kind of the I-V curves leading to a 'gap map'. The STS data can be correlated to the STM topography data for visualisation of the local electronic structure within one grain.

Figure 1 shows the topographic STM images with a scan size of 500×500nm$^2$ for the LSMO and LCMO films. The scans were recorded at sample temperatures of 87K and 96K, respectively. The sputtered LSMO film exhibits a microstructure with an average grain size of 46 nm. The root mean square (RMS) roughness is about 2.1 nm. The LCMO film, however, shows a larger average grain size of 57 nm and a slightly larger RMS roughness of 2.9 nm. It consists of spherically shaped dense packed grains with a small size distribution, whereas the LSMO film shows irregularly shaped grains with a broad size distribution. The difference in the grain shape may be attributed to the sputtering (LSMO) and MAD (LCMO) preparation techniques.

In Fig. 2b-2d the conductance maps of the LSMO sample at different measurement temperatures of 87K, 150K, and 278K are shown. First the discrete array of conductance sampling points was smoothed using a Gaussian averaging algorithm and then a threshold criterion of 5.9×10$^{-3}$ nA/V (see discussion of Fig. 4) was set for generating a black and white 2 dimensional structure. Insulating areas are shown in black while white regions correspond to metallic behavior. At 87K insulating filament like domains are observed. The average width of the filaments is about 21nm which is well below the observed grain size. This result shows that the domain structure is not directly correlated to the microstructure and therefore can be attributed to intrinsic effects. With increasing temperature a network of these insulating regions builds up, leaving a few metallic islands within this mesh. The number and size of the conductive domains continuously decreases up to a temperature of 278K. This scenario can be compared with the percolation model of Burgy et al [13] where also insulating filaments were proposed. The observed domain structure is much finer than that shown by Fäth et al. [15] where the contrast at a voltage of 3V was used.

Figure 3 combines the topographic information with the conductance map of the LCMO surface to a three dimensional (3D) picture. Red areas represent high conductivity zones (max. 0.021 nA/V) while black regions are insulating. As found in measurements on LSMO, the domain structure is not directly correlated with the grains. The gap width evaluated from the STS measurements for LCMO and LSMO films is temperature independent with an average value of 0.15 eV.

To visualize the different phases as a function of temperature in Fig. 2 a threshold of $5.9 \times 10^{-3}$ nA/V for the conductance was chosen which distinguishes between insulating (black) and metallic regions (white). This threshold value was adjusted with the help of the magnetization curve: the magnetization at low temperatures was assumed to be proportional to the total area of the conducting domains, and the threshold chosen so as to reproduce the measured magnetization of the sample. As one can see in Fig. 4 the agreement between STS and magnetization measurement using a single threshold criterion is very good. Deviations near $T_c$ are probably due to the temperature dependence of the magnetization of the individual conducting domains, which was neglected.

The mean tunneling conductance of the metallic regions as a function of temperature could also be evaluated from those conductance values above this threshold. The tunneling resistance (circles in Fig. 4) increases slightly with increasing temperature but cannot explain the shape of the MIT since only the metallic regions are taken into account, and current must also start to flow through insulating domains as these percolate. We note that the absolute values of the transport resistance cannot directly be compared with those of the tunneling resistance, because of the very different character of the measurements. Unfortunately, due to the high conductance near $T_c$, we are not able to determine the temperature dependence of the conductance of the insulating regions in order to fully model the MIT according to the percolation transition.

To support the interpretation we performed simulations in terms of a resistor network: The measured local tunneling conductance from STS experiments was used to create a resistor network and the resistance of the whole network is given in Fig. 4 as a function of temperature. The simulation shows clearly the development of current channels through the sample near the transition temperature as expected from the proposed percolation behavior. The shape of the simulated transition corresponds with the transport data although our data only represent a small part of the sample leading to deviations from the transport measurements especially in the high temperature regime.

Without intrinsic inhomogeneities one would expect that the resistance jumps from the metallic to the insulating resistance branch at the Curie temperature. This is not the case in real samples: The sample falls into metallic and insulating regions. The development of the insulating area with temperature shows the influence of the geometry of the domains on the resistance as a function of temperature and explains the broad transition seen in the resistance.

In our STS investigations the influence of growth-induced inhomogeneities or surface contaminations cannot be excluded because the samples had been exposed to ambient air between preparation and STS measurements. However, stable topographic images and the development of the electronic phase separation with temperature (which is reversible!) proves that the intrinsic behavior shown in our experiments cannot be dominated by such artefacts.

In summary, we have shown that the MIT in manganite thin films is dominated by intrinsic inhomogeneities. The metallic and insulating regions can be visualized with the STS technique. Our measurements show clearly that the broad transition in the resistance of LSMO thin films can be attributed to the spatial distribution of the different phases in the nanometer range and therefore support the theoretically predicted percolation behavior. The visualized domain structure is not correlated to the microstructure and is an intrinsic phenomenon of the samples.

**Acknowledgements**

The work was supported by Deutsche Forschungsgemeinschaft, SFB 602, TPA2.

**References**

1.) G.H. Jonker, J.H. van Santen, Physica **16**, 337 (1950)

2.) R. von Helmolt, *et al*., Phys. Rev. Lett. **71**, 2331 (1993)

3.) M. Imada, A. Fujimori, Y. Tokura, Rev. Mod. Phys. **70**, 1039 (1998)

4.) Y. Tokura, Gordon, *Colossal magnetoresistive oxides*, (Breach Science Publishers, Amsterdam, 2000)

5.) T.A. Kaplan, S.D. Mahanti, *Physics of Manganites*, (Kluwer Academic/Plenum Publishers, New York,1999)

6.) C. Zener, Phys. Rev. **82**, 403 (1951)

7.) P.W. Anderson, H. Hasegawa, Phys. Rev. **100**,675 (1955)

8.) A. Urushibara *et al.,* Phys. Rev. *B* **51**. 14103 (1995)

9.) P. Schiffer, A.P. Ramirez, W. Bao, S.-W. Cheong, Phys. Rev. Lett. **75**, 3336 (1995)

10.) S. Yunoki, A. Moreo, E. Dagotto, Phys. Rev. Lett. **81**, 5612 (1998)

11.) A. Moreo, M. Mayr, A. Feiguin, S. Yunoki, E. Dagotto, Phys. Rev. Lett. **84**, 5568 (2000)

12.) M. Mayr *et al.* , Phys. Rev. Lett. **86**, 135 (2001)

13.) J. Burgy, M. Mayr, V. Martin-Mayor, A. Moreo, E. Dagotto, Phys. Rev. Lett. **87**, 277202 (2001)

14.) M. Uehara, S. Mori, C.H. Chen, S.-W. Cheong, Nature **399**, 560 (1999)

15.) M. Fäth *et al.,* Science **285**, 1540 (1999)

16.) Q. Lu, C.-C. Chen, A. de Lozanne, Science **276**, 2006 (1997)

17.) Ch. Renner *et al.*, Nature **416**, 518 (2002)

18.)  Y. Luo, A. Käufler, K. Samwer, Appl. Phys. Lett. **77**, 1508 (2000)

19.)  Y. Luo, K. Samwer, J. Appl. Phys. **89**, 6760 (2001)

20.)  V. Moshnyaga *et al*., Appl. Phys. Lett. **74**, 2842 (1999)

21.)   V. Moshnyaga *et al.,* J. Appl. Phys. **86**, 5642 (1999)

**Figure captions**

Fig. 1. Topographic STM images for LSMO and LCMO, 500×500nm², at 87K (LSMO) and 96K (LCMO)

Fig. 2. A: Two different I-U characteristics for LSMO in metallic and insulating regions; B-D: Conductance maps, 500×500nm², black: insulating regions (dI/dV<5.9×10$^{-3}$nA/V), white: metallic regions (dI/dV>5.9×10$^{-3}$nA/V) for 87K, 150K and 278K. Area of insulating domains grows with increasing temperature.

Fig. 3. Three dimensional image of LCMO, 170x170nm$^2$, at 186K. The topographic STM image is shown with the coloring of the conductance map, black: insulating regions, red: conductive regions (max 0.021nA/V)

Fig. 4. Electrical resistance (full line, transport), normalized tunneling resistance (circles, STS) and normalized network resistance (stars, simulation) as a function of temperature for LSMO (left scale). Additionally the magnetization curve (dotted line) together with the relative part of the area of the metallic domains (squares) is shown.

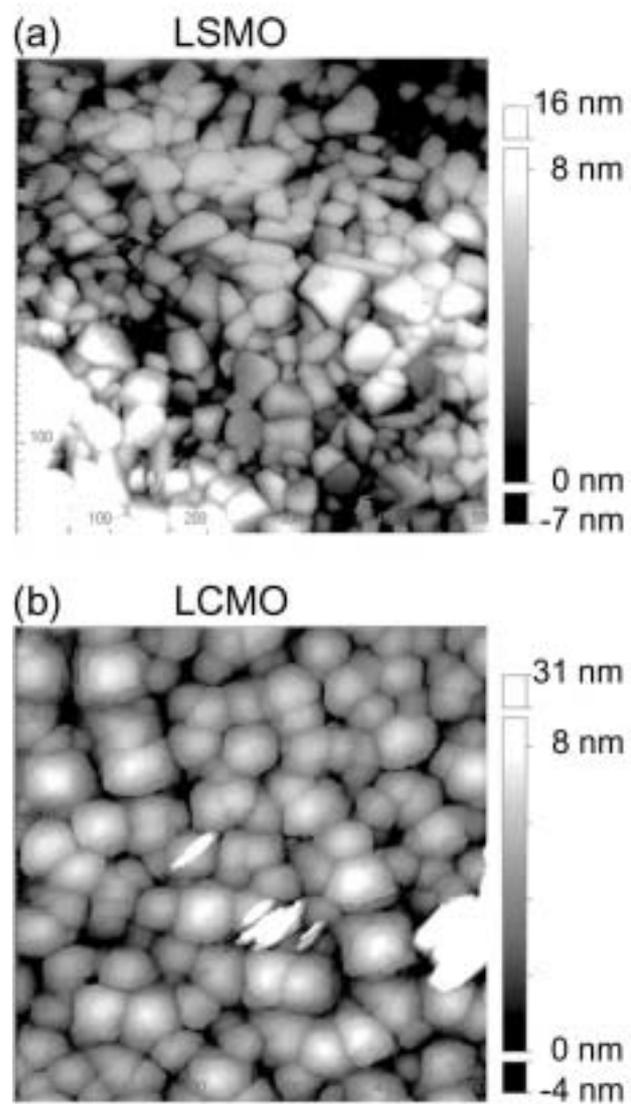

Fig. 1: Th. Becker et al

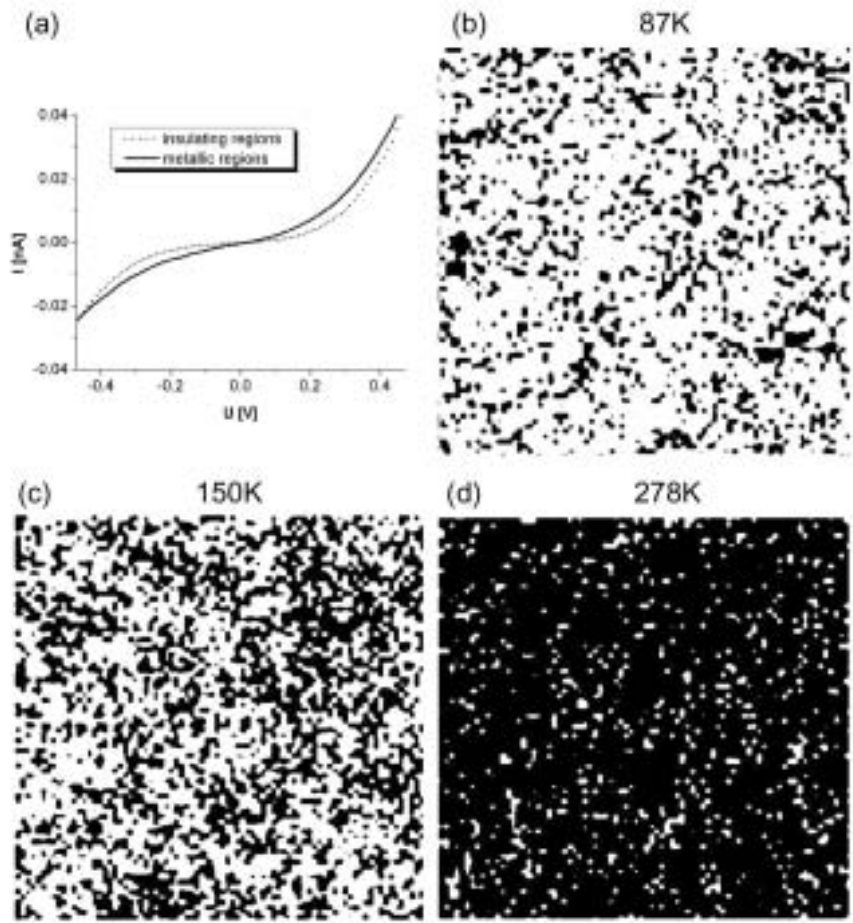

Fig. 2: Th. Becker et al

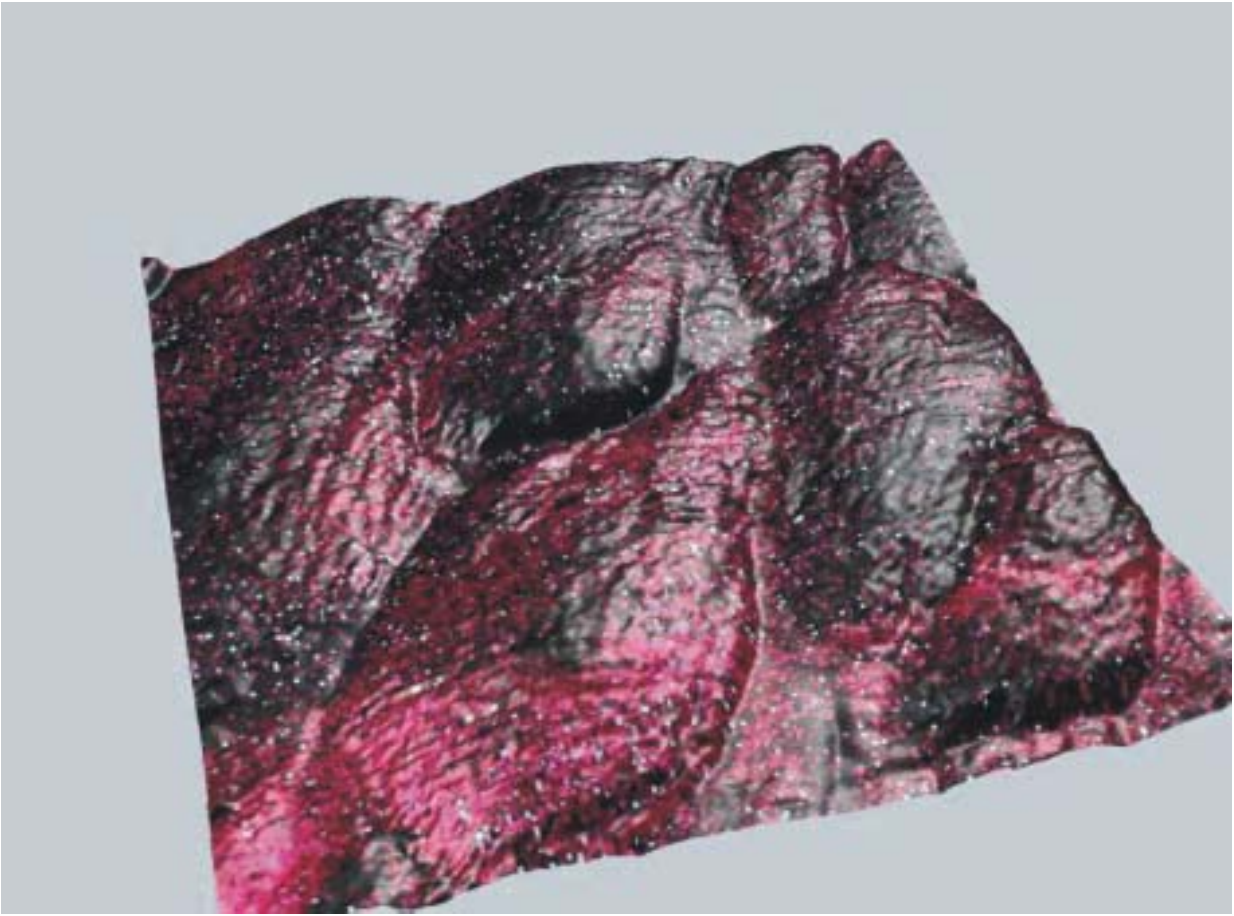

Fig. 3: Th. Becker et al

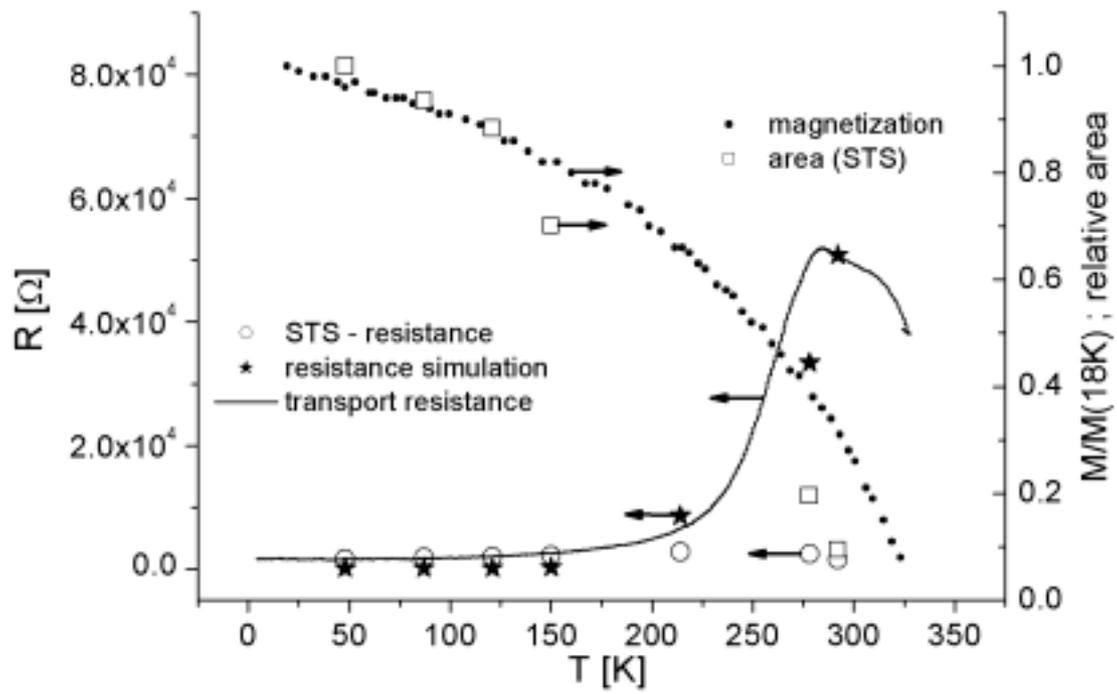

Fig. 4: Th. Becker et al